\documentclass[aps,prb,showpacs,twocolumn,superscriptaddress]{revtex4}
\usepackage{graphicx}
\usepackage{bm}
\usepackage{amsmath}

\def\be{\begin{equation}}
\def\ee{\end{equation}}
\def\bea{\begin{eqnarray}}
\def\eea{\end{eqnarray}}

\begin{document}

\title{Spin Hall effect of conserved current: Conditions for a nonzero spin Hall current}

\author{Naoyuki Sugimoto}
\email[Electronic address:]{sugimoto@appi.t.u-tokyo.ac.jp}
\affiliation{Department of Applied Physics, University of Tokyo, Hongo,
Bunkyo-ku, Tokyo 113-8656, Japan}

\author{Shigeki Onoda}
\affiliation{Spin Superstructure Project, ERATO, Japan Science and
Technology Agency, c/o Department of Applied Physics, University of
Tokyo, Hongo, Bunkyo-ku, Tokyo 113-8656, Japan}

\author{Shuichi Murakami}
\affiliation{Department of Applied Physics, University of Tokyo, Hongo,
Bunkyo-ku, Tokyo 113-8656, Japan}

\author{Naoto Nagaosa}
\affiliation{Department of Applied Physics, University of Tokyo, Hongo,
Bunkyo-ku, Tokyo 113-8656, Japan}
\affiliation{CERC, AIST,Tsukuba Central 4, Tsukuba 305-8562, Japan}
\affiliation{CREST,Japan Science and Technology Agency (JST)}

\begin{abstract}
We study the spin Hall effect taking into account the impurity 
scattering effect as general as possible with the focus on the 
definition of the spin current.
The conserved bulk spin current (Shi {\it et al.} [Phys. Rev. Lett. {\bf 96}, 076604 (2006)] ) 
satisfying the continuity equation of spin is 
considered in addition to the conventional one defined by 
the symmetric product of the spin and velocity operators. 
Conditions for non-zero spin Hall current are clarified. 
In particular, it is found that
(i) the spin Hall current is non-zero in the Rashba model with a
finite-range 
impurity potential, and (ii) the spin Hall current vanishes in the 
cubic Rashba model with a $\delta$-function impurity potential.  
\end{abstract}
\maketitle

Spintronics is one of the most promising new technologies, where the 
spin degrees of freedom of electrons in semiconductors are manipulated 
and utilized for functions such as memory, operation, and 
communication.~\cite{review,Ohno} One of the key routes to spintronics 
is to invent an efficient method to inject spins into semiconductors.
In this respect, the intrinsic spin Hall effect (SHE) 
has attracted recent intensive attention since its 
theoretical proposal.~\cite{Murakamiscience, Sinovaprl} This can give 
a larger effect by orders of magnitude than the extrinsic one based 
on the impurity scatterings proposed long 
before.~\cite{Dyakonov,Hirsch,Zhang}  

Recently two experiments have been reported on observations of the
SHE in GaAs and related materials.\cite{Kato,Wunderlich} 
Kato {\it et al.}\cite{Kato} observed the Kerr rotation due to the spin
accumulation 
($\sim 10 \mu_{B} (\mu \mathrm{m})^{-3}$) near the edges of the n-type
GaAs sample. 
They suggested the extrinsic mechanism of the SHE since 
it was almost insensitive to the crystal orientation.
Wunderlich {\it et al.}\cite{Wunderlich} observed the circularly polarized
LED signal from spin-polarized interfacial two-dimensional holes in p-type 
GaAs system. From an estimation of transport lifetime, they concluded that 
the spin accumulation is due to the intrinsic SHE.

The debates on the impurity effect on the intrinsic 
SHE have continued, which are in parallel to those for the anomalous 
Hall effect in ferromagnets. In the latter case, the intrinsic mechanism of 
Karplus-Luttinger~\cite{KarplusLuttinger} was criticized and extrinsic 
mechanisms due to impurity scatterings were 
proposed.~\cite{Smit} The Hall conductivity is a singular 
function as the disorder strength approaches zero. 
In the metallic case, the vertex correction in the diagrammatic 
language incorporates a deviation of the electronic distribution
function from equilibrium, and it represents the dissipative current. 
This situation is similar also in the spin Hall current.

Actually, the disorder effect on the spin Hall current of the Rashba model 
in two dimensions has been intensively studied.
Sinova {\it et al.}~\cite{Sinovaprl} 
obtained the universal value $e/(8 \pi)$ \cite{note8pi}
for the spin Hall conductivity (SHC) 
$\sigma_H^s$ without disorder. When the 
self-energy correction due to impurity 
scattering is taken into account, 
the SHC $\sigma_H^s$ 
is reduced continuously as a function of the disorder strength from the 
universal value $e/(8\pi)$.~\cite{Loss} 
On the other hand, Inoue {\it et al.}~\cite{Inoue} 
studied the vertex correction and found that
$\sigma^s_H$ vanishes in the clean limit within the Born approximation.
Furthermore, it has been shown that the
SHC $\sigma^{s}_{H}$ vanishes for any value of the
lifetime $\tau$ using the  Keldysh formalism,~\cite{Halperin,Liu,Liu2} 
the Kubo formula analytically,~\cite{Dimitrova,Chalaev,Raimondi} 
and numerically,~\cite{Sheng,Nomura2} and the Boltzmann equation.~\cite{Khaetskii} 
Thus after long debates, people have reached the consensus that
the SHC $\sigma^{s}_{H}$ for the Rashba model
vanishes for any $\tau$.~\cite{previous} 

However, the vanishing result of $\sigma^{s}_{H}$
depends on the definition of the spin current.
In the previous calculations on the SHC, the spin 
current is defined {\it ad hoc} as a symmetrized product of 
the spin and the velocity $J_{s}=\frac{1}{2}\{ v_{y},s_{z}\}$,
where $v_{y}=\partial H/\partial p_{y}$,
in response to the electric field $\mathbf{E}$ along the $x$ axis. 
However, this ``conventional'' definition of spin current loses its
physical foundation when the spin-orbit coupling is present and the 
conservation law of the spin is violated.
This is an important issue since the concept of ``current'' depends crucially 
on conservation; non-local effects of the current comes from 
the fact that an incoming flow goes out without loss. Therefore we 
need to search for a 
proper definition of a conserved spin current in the bulk. 
From this viewpoint, the conserved spin current ${\cal{\bm J}}_{s}$~\cite{PZhang} 
deserves scrutiny. 
If $\int dV\langle \dot s_z\rangle=0$, 
it satisfies the Onsager's reciprocity relation and the continuity equation for the spin in the bulk: 
$\partial_t \langle s_z\rangle+\nabla \cdot{\vec{\cal J}}_s=0$. 
Furthermore it is free from an artifact that the spin current is
proportional to time derivative of the spin operator;~\cite{Dimitrova,Chalaev} 
hence, the spin Hall current can 
be nonzero even for the Rashba model. 
Even though the experiments on the SHE up to now \cite{Kato,Wunderlich}
detect the spin accumulation at the 
sample edges, we stick here to the SHE defined by the bulk spin current.
This is because we consider that 
the generation of the conserved spin current in the bulk is 
a more fundamental phenomenon than the spin accumulation 
at edges, which is not solely determined by this spin current since
the continuity equation for the spin is not satisfied there. 
In principle, there should be other means to detect the SHE without using the 
spin accumulation such as voltage measurement with the injected spin current.~\cite{spininjection}

In this paper, we study the SHE as generally as possible taking into account
the disorder 
with the definitions of the 
conventional and conserved spin current. 
Applying this consideration, some new results are obtained for the 
Rashba and cubic Rashba models. 
We employ the Keldysh formalism,~\cite{Rammer,Halperin,mahan} by which 
the infinite series of the Feynman diagrams both 
for the self-energy and vertex correction are taken into 
account compactly, and the expression for the SHC 
is obtained for both definitions of the spin current 
described above.

We consider a generic model with spin-orbit coupling 
with a random impurity potential, 
this random potential is assumed to be spin-independent, 
and the time-reversal symmetric model 
exclusively. 
In the Keldysh formalism, a Green's function matrix $\underline{G}$ 
is introduced, 
\begin{equation}
\underline{G}=\left(\begin{array}{cc}
G^{R}&G^{<}\\
0 & G^{A}
\end{array}
\right),
\end{equation}
where the superscripts $R$, $A$, and $<$ denote 
the retarded, advanced and lesser Green's functions, respectively.
The self-energy matrix $\underline{\Sigma}$ is defined similarly.
The Green's functions satisfy
\begin{eqnarray}
(G^R)^{-1}\otimes G^<
-G^<\otimes(G^A)^{-1}
&=&\Sigma^< \otimes G^A
-G^R\otimes\Sigma^<,\nonumber\\
\\
({\cal G}_0^{-1}-\Sigma ^{R,A})\otimes G^{R,A}&=&\delta (1-2),
\end{eqnarray}
where 
$(A\otimes B)(1,2)\equiv \int d3 A(1,3)B(3,2)$ and 
${\cal G}_0$ is the unperturbed Green's function. 
We then separate the center-of-mass and the relative
coordinates and perform the Fourier transform to the 
relative coordinates.~\cite{mahan,Rammer} 
The final result is written in terms of 
the center-of-mass coordinates $(T,\mathbf{R})$ and 
the relative momentum $(\omega,\mathbf{p})$.
We put the constant electric field $\mathbf{E}=(E,0,0)$ 
and look for solutions independent of $T$ and $\mathbf{R}$. 
Therefor the quantum Boltzmann
equation (QBE) is written by 
\begin{eqnarray}
&&[{\underline G},~H]=
-ie{\mathbf E}\cdot{\mathbf{\nabla_p}}{\underline G}
-\frac{i}{2}e{\mathbf E}\cdot
\left\{{\mathbf\nabla_p} H,~
\partial_\omega{\underline G}\right\}\nonumber\\
&&-\frac{i}{2}e{\mathbf E}\cdot\left(
\left\{{\mathbf\nabla_p} {\underline\Sigma},~
\partial_\omega{\underline G}\right\}
-\left\{\partial_\omega {\underline \Sigma},~
{\mathbf\nabla_p}{\underline G}\right\}
\right)
+[{\underline\Sigma},~{\underline G}]
.\label{QBE0}
\end{eqnarray}

\begin{figure}[t]
\includegraphics[width=5cm,clip]{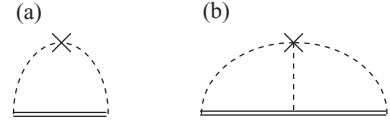}
\caption{\label{fig:diagram} Diagrammatic representation of the self-energy 
$\underline{\Sigma}$ in 
the present self-consistent approximation. Doubled line is the dressed 
Keldysh Green's function $\underline{G}$. Dashed line with a cross denotes the average over 
the impurity positions in the second order for (a) and in the third 
order (b), respectively.}
\end{figure}

Here one can show from Eq.~(\ref{QBE0}) that the 
time derivative $\dot{{\cal O}}$  of an arbitrary 
operator ${\cal O}$, which is independent of $\mathbf{p}$, $\mathbf{R}$, 
$T$ and $\omega$, has a vanishing expectation value in the steady state, 
$\langle \dot{{\cal O}}\rangle =0$, 
even with a general form of the impurity potential.~\cite{Dimitrova,Chalaev} 
We start with 
\begin{eqnarray}
&&\langle \dot{ {\cal O}} \rangle =
\frac{1}{i}\int\frac{d\omega}{2\pi}
\int\frac{d^{2}p}{(2\pi)^{2}} \mathrm{tr}(\dot{ {\cal O}} G^{<})
\label{Js0} \nonumber \\
&&\ \ \ =
\int\frac{d\omega}{2\pi}
\int\frac{d^{2}p}{(2\pi)^{2}} \mathrm{tr}({\cal O}[{\underline G},H]^<),
\label{Js}
\end{eqnarray}
where $[\ \ ]^{<}$ denotes the lesser (upper right) component 
of the matrix in the Keldysh space. 
Eq.~(\ref{QBE0}) is plugged into Eq.~(\ref{Js}), and evaluate 
the respective terms in the RHS of Eq.~(\ref{QBE0}). 
The first term $ie\mathbf{E}\cdot{\mathbf\nabla_p}G^{<}$ becomes zero
after an integration over $\mathbf{p}$. 
The second term vanishes after an $\omega$-integration.
The third and fourth terms vanish after partial integrations 
in terms of $\mathbf{p}$ and $\mathbf{\omega}$. 
Lastly, to evaluate  
the last term $([\underline{\Sigma},\ \underline{G}]^{<})$, 
we need a relationship between the self-energies and Green's functions. 
We employ the self-consistent Born approximation [the diagrams in 
Fig.~\ref{fig:diagram}~(a) and (b)]~\cite{Crepieux} 
for the impurity scattering. Up to the second order it is given by 
\begin{eqnarray}
&&\underline{\Sigma}(\omega,\mathbf{p})=n_{i}\int\frac{d^{2}p'}{(2\pi)^{2}}
\left|V_{\mathbf{p},\mathbf{p}'}\right|^{2}\underline{G}(\omega,\mathbf{p}')
\nonumber \\
&&
+n_{i}\int\frac{d^{2}p'\, d^{2}p''}{(2\pi)^{4}}
V_{\mathbf{p},\mathbf{p}'}V_{\mathbf{p}',\mathbf{p}''}
V_{\mathbf{p}'',\mathbf{p}}
\underline{G}(\omega,\mathbf{p}')
\underline{G}(\omega,\mathbf{p}'').\ \ \ 
\label{2-Born}
\end{eqnarray}
where $n_{i}$ is an impurity density, and $ V_{\mathbf{p},\mathbf{p}'}$ 
is the Fourier transform 
of the impurity potential. 
From the second Born approximation Eq.~(\ref{2-Born}) one can easily 
show $\langle \dot{\cal O} \rangle =0$ for arbitrary forms of 
impurity potentials\cite{Note}. 
This holds true even for higher-order Born approximation. 
On the other hand, 
for the charge current $\mathbf{v}=\dot{\mathbf{R}}$, this 
argument does not apply, and $\langle \mathbf{v} \rangle$ can be nonzero 
in the steady state as expected.

Now we proceed to a closed form of the SHC.
The expectation value of the conventional spin current 
$J_{s}\equiv \frac{1}{2}\{v_y,s_z \}$ 
is obtained as 
\begin{eqnarray}
\frac{\langle J_s \rangle}{E}  &=& 
-\lim_{E\rightarrow 0}\frac{i}{2E}\int\frac{d\omega}{2\pi}
\frac{d^{2}p}{(2\pi)^{2}}
\mathrm{tr}
\left[s_{z}
\left\{\frac{\partial H}{\partial p_{y}},\underline{G}\right\}
\right]_{E}^<,
\end{eqnarray}
where $[\ ]_E$ implies that we retain the terms  linear in a 
uniform electric 
field $\mathbf{E}=E\hat x$. 

Let us turn to the second definition of the spin current
${\vec {\cal J}}_{s}$ as proposed by Shi {\it et al}.~\cite{PZhang} 
${\vec {\cal J}}_{s}$ is defined to satisfy 
$\partial_t \langle s_z\rangle+\nabla\cdot{\vec {\cal J}}_s=0$, 
and is divided into $\langle {\vec J}_s\rangle+{\vec P}_{\tau}$, 
where ${\vec J}_s=\frac{1}{2}\{{\vec v},s_z\}$. 
The second term ${\vec P}_{\tau}$ is called the torque dipole density, and 
is required to satisfy $\lim_{\Omega\rightarrow 0}\lim_{{\vec Q}\rightarrow 0}
(\langle \dot s_z(\Omega,\vec{Q})\rangle+i\vec{Q}\cdot\vec{P}_{\tau}(\Omega,{\vec Q}))=0$, 
where ($\Omega,{\vec Q}$) are the Fourier components 
of the center-of-mass coordinates. 
We put $Q_x=0$, and take the limit $Q_y=Q \rightarrow 0$, i.e., 
${P}_{\tau}\equiv{\vec P}_{\tau}\cdot\hat{\mathbf y}=
-\lim_{\Omega\rightarrow 0}
\lim_{Q\rightarrow 0}\frac{1}{iQ}
\langle\dot{s}_{z}(\Omega,Q)\rangle$. 
More explicitly, $\mathbf E$ is spatially modulated along the y-axis, 
$\mathbf{E}=Ee^{iQY-i\Omega T}\hat{\mathbf{x}}$.~\cite{PZhang} 
Note that $\langle\dot{s}_z(0,0)
\rangle=0$ from Eq.~(\ref{Js}), which 
means $P_{\tau}$ is finite and well-defined. 
In response to the electric field, 
the Green's function $\underline{G}$ and the self-energy $\underline{\Sigma}$
acquire terms proportional to 
$e^{iQY-i\Omega T}$. 
$P_{\tau}$ is expressed as
\begin{eqnarray}
&&P_\tau=
-i\lim_{Q\rightarrow 0}
\frac{1}{Q}\int\frac{d\omega}{2\pi}
\int\frac{d^{2}p}{(2\pi)^{2}} \mathrm{tr}(s_{z}[H, \underline{G}]^<).
\label{Ps1}
\end{eqnarray}
We first write down the QBE to the linear order in 
$E$ and to the linear order in $\Omega$ or $Q$.  
Next, we replace the term $[H,\underline{G}]^<$ in Eq.~(\ref{Ps1})
with the corresponding term in the QBE. 
While there arise a number of terms, most of them give no contribution to 
$P_{\tau}$
after partial integrations over $p_i$ or $\omega$.
To calculate the remaining terms, we 
note that this electric field necessarily accompanies a magnetic field 
according to the Maxwell equation.
To deal with the response to these two fields, it is convenient to consider the
corresponding vector potential,
$\mathbf{A}=(Ae^{iQY-i\Omega T},0,0)$. 
Relevant terms in the QBE are classified to 
those proportional to $i\Omega A=E$ (electric field) or
those proportional to $-iQA=B$ (magnetic field).
The resulting form is
a sum of the contributions from 
the response to a dc electric field $E\hat{x}$ and 
that to a dc magnetic field $B\hat{z}$:
\begin{eqnarray}
&&\frac{{\cal J}_{s}}{E}
=\lim_{E\rightarrow 0}\frac{i}{E}\int\frac{d\omega}{2\pi}
\frac{d^{2}p}{(2\pi)^{2}} \mathrm{tr}
\left[\frac{1}{2}s_z\left\{
\frac{\partial \underline{\Sigma}}{\partial p_{y}},\ \underline{G}
\right\}\right]^{<}_{E}\nonumber \\
&&+\lim_{B\rightarrow 0}
\frac{i}{B}
\int\frac{d\omega}{2\pi}
\frac{d^{2}p}{(2\pi)^{2}}\mathrm{tr}
\left[s_z
\underline{G}\!-
\frac{1}{2}s_z\left\{
\frac{\partial\underline{\Sigma}}{\partial \omega},\,\underline{G}
\right\}\right]^{<}_{B}.
\label{eq:calJ}
\end{eqnarray}
Here ${\cal J}_s={\vec {\cal J}}_s\cdot\hat{\mathbf y}$, and $[\ ]_{B}$ retains the terms linear in an uniform 
magnetic field $\mathbf{B}=B\hat{\mathbf z}$. 
Because $\mathbf{B}$ does not drive the system off-equilibrium,
the relation in equilibrium, $G^<_0=(1-\tanh(\omega/{2k_BT}))(G^A_0-G^R_0)/2$, 
is satisfied even in the presence of $\mathbf{B}=B\hat{\mathbf z}$. 
The details of the derivation of Eq.~(\ref{eq:calJ}) will be 
presented elsewhere.~\cite{Sugimotolong}. 
Remarkably, in calculating the total conserved spin current 
${\cal J}_{s}=\langle J_s\rangle +P_{\tau}$,
there appears a term in $P_{\tau}$ 
which exactly cancels $\langle J_s\rangle$.
We also note that this formula is quite generic and 
applies to any models. The expression of the charge current
is obtained just by replacing $s_z$ by $-e$ in Eq.~(\ref{eq:calJ}), 
and the $B$-term in the final formula is reminiscent of 
the Streda formula.~\cite{Streda1,Streda2} 

We now discuss the explicit models based on the results obtained above.
Here we condider the Rashba model
\begin{equation}
H=\frac{{\bf p}^2}{2m} + \lambda ({\bm\sigma}\times\mathbf{p}) \cdot 
 {\hat z}  + v,
\label{Rashba}
\end{equation}
and the cubic Rashba model
\begin{equation}
H=
\frac{{\bf p}^2}{2m} + \frac{i\lambda}{2} (p_{-}^{3}\sigma_{+}
-p_{+}^{3}\sigma_{-})
  + v.
\end{equation}
Here, ${\bm\sigma}=(\sigma_x,\sigma_y,\sigma_z)=2\mathbf{s}$ 
is the Pauli matrix, 
$p_{\pm}=p_x\pm ip_y$, $\sigma_{\pm}=\sigma_{x}\pm i\sigma_{y}$
and $v$ is an impurity random potential.
We take the unit where $\hbar=c=1$. 
The Rashba model represents an n-type semiconductor in two-dimensional 
heterostructure. The second term in Eq.~(\ref{Rashba}) represents 
the spin-orbit coupling with an inversion-symmetry-breaking
potential along the $z$ direction perpendicular to the plane.
As noticed in Refs. \onlinecite{Dimitrova} and \onlinecite{Chalaev},
$J_s$ is proportional to $\dot{s}_{y}$: 
$J_s=\frac{p_y\sigma_z}{2m}=\frac{[H,\sigma_{y}]}{4im\lambda}$.
Therefore from the generic argument in Eq.~(\ref{Js}), there occurs no
spin Hall current when this definition of the spin current is employed for
the Rashba model (Table I(a)). 
This result is consistent with previous works using various methods, 
including the calculations in the clean limit 
by Kubo formula,~\cite{Inoue,Raimondi} and Keldysh formalism.~\cite{Liu} 
For finite $\tau$ $(\epsilon _F\tau\gg 1)$, it is also consistent with the 
analytic~\cite{Dimitrova,Chalaev} and numerical~\cite{Sheng,Nomura2} 
results
by Kubo formula, 
and with the results by Keldysh formalism.~\cite{Halperin,Liu2} 
The calculation in Ref.~\cite{Chalaev} 
by Kubo formula for finite $\tau$ $(\epsilon _F\tau\gg 1)$ 
is similar to ours by the Keldysh formalism. 
Nevertheless, our approach better reveals the reason why 
the spin Hall current generally vanishes when $J_s\propto{\dot s}_y$. 

The cubic Rashba model, on the other hand, describes the heavy-hole 
bands of cubic semiconductors in heterostructute.
The conventional spin current $J_s$ can no longer be expressed as
$\dot{\cal O}$; hence the previous argument for vanishing spin Hall current 
 does not apply. Indeed, the resulting SHC
is nonzero \cite{BZn} [Table I(b) ], 
which is consistent with Refs.~\onlinecite{holegas1} and \onlinecite{holegas2}. 
Therefore the zero or nonzero spin Hall current for the 
conventional spin current $J_s$ is mostly
determined whether it is expressed by the time derivative of some
local operator or not. 

Next we turn to the conserved spin current ${\cal J}_s$. 
Both in the Rashba and the cubic Rashba models,
the $B$-term in Eq.~(\ref{eq:calJ}) vanishes,
because the Hamiltonian lacks a $\sigma_z$-term, and the 
self-energy is independent of the spin.
The remaining $E$-term in Eq.~(\ref{eq:calJ}) for 
the Rashba model is calculated as follows.
In the first Born approximation [the first term in Eq.~(\ref{2-Born})], 
${\cal J}_s=0$ for general $V_{\mathbf{p}-\mathbf{p}'}$. 
In the second Born approximation,
we obtain
\begin{eqnarray}
&&{\cal J}_s=
-\frac{in_i}{4}\int\frac{d\omega}{2\pi}
\frac{d^{2}p}{(2\pi)^{2}} 
\frac{d^{2}p'}{(2\pi)^{2}} 
\frac{d^{2}p''}{(2\pi)^{2}} 
\mathrm{tr}\Big[\sigma_{z}
\big[
\underline{G}(\omega,\mathbf{p})
\nonumber \\
&&\ \
\cdot
\underline{G}(\omega,\mathbf{p}')
\underline{G}(\omega,\mathbf{p}'')
\big]_E^<
\frac{\partial (V_{\mathbf{p}-\mathbf{p}'}
V_{\mathbf{p}'-\mathbf{p}''}V_{\mathbf{p}''-\mathbf{p}})}{\partial p'_{y}}
\Big].
\end{eqnarray}
Even for higher-order Born approximation, the formula for 
${\cal J}_{s}$ can be written down. We can then 
see that ${\cal J}_{s}$ always depends on $\partial_{p_{y}}V$,
and the spin Hall current for ${\cal J}_{s}$ is extrinsic for both the 
Rashba and the cubic Rashba models, 
depending explicitly on the impurity potential.
All these considerations are summarized in Table I. 
With the $\delta$-impurity potential the conserved 
spin Hall current vanish both for the Rashba model and the cubic Rashba model.  
We note that calculated the conserved SHC without disorder 
is $\sigma^s_H=e/8\pi$ for the Rashba model and 
$\sigma^s_H=-9e/8\pi$ in the cubic Rashba model.~\cite{PZhang} 
These results are reproduced in our calculations by neglecting the self-energy of 
the lesser Green's function and taking the clean limit.

\begin{table}
(a) Rashba model   \begin{tabular}{c|c|c|c}\hline\hline
      Impurity potential          & Born approx. &\multicolumn{2}{c}{Definition of spin current}\\
      \cline{3-4}
                                  &  & \ \ \ \ \ \ $\langle J_s\rangle$ \ \ \ \ \ \ & ${\cal{J}}_s$\\
      \hline
      $\delta(\mathbf{r})$        & 1st/higher & 0 \cite{IK} & \bf{0}\\

      \hline
      $V_{\mathbf{p}-\mathbf{p}'}$ & 1st    & 0 \cite{long} & \bf{0}\\
      \cline{2-4}
                                  & higher & \bf{0} & \bf{Finite}\\
                                  \hline\hline
    \end{tabular}

\vspace{1mm}

(b)\ Cubic Rashba model

    \begin{tabular}{c|c|c|c}\hline\hline
      Impurity potential          & Born approx. &\multicolumn{2}{c}{Definition of spin current}\\
      \cline{3-4}
                                  &  & \ \ \ \ \ \ $\langle J_s\rangle$ \ \ \ \ \ \ & ${\cal J}_s$\\
      \hline
      $\delta(\mathbf{r})$        & 1st/higher & Finite \cite{holegas1,holegas2} & \bf{0}\\
      \hline
      $V_{\mathbf{p}-\mathbf{p}'}$ & 1st    & Finite & \bf{0}\\
      \cline{2-4}
                                  & higher & Finite & \bf{Finite}\\
                                  \hline\hline
    \end{tabular}
  \caption{\label{table}Spin Hall effect in the (a) (linear) 
Rashba model and (b) cubic Rashba model, with various types of the 
impurity potential for the two different definitions: 
$\langle J_s\rangle$ is the conventional spin current, 
$J_s=\frac{1}{2}\{v_y,s_z\}$, and ${\cal{J}}_{s}$ is the 
conserved effective spin current, 
${\cal{J}}_s=\langle J_s\rangle +P_{\tau}$.
We show the new results in the boldface.}
\end{table}

\begin{figure}[t]
\includegraphics[width=6.5cm,clip]{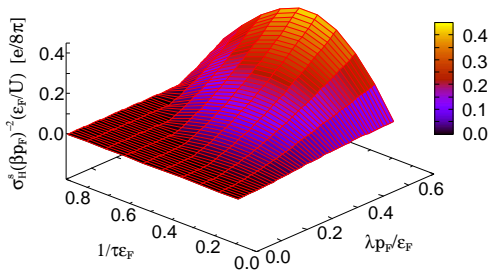}
\caption{\label{fig:hd} Contour map of the SHC 
$\sigma^s_H$. It has the maximum value 
nearly along $\alpha = 2\lambda p_F \tau =1$.
The momentum cutoff is set to be $\Lambda\approx 3 p_{F}$, and 
$U/\varepsilon_F=0.02$.}
\end{figure}

We now demonstrate the finite spin Hall current for ${\cal{J}}_s$ 
in the Rashba model with the short-range (not $\delta$-function)
impurity potential $V(r)=Ue^{-(\frac{r}{\beta})^2}$ with $\beta\Lambda\ll 1$, 
where $U$ is the magnitude of the impurity potential,
$\beta$ is the size of the potential range and $\Lambda$ is a momentum cutoff. 
As an approximation, we substitute the Green's functions in 
Eq.~(\ref{eq:calJ}) with those for the 
$\delta$-function impurity potential,
and calculate ${\cal J}_{s}$.
Figure \ref{fig:hd}
shows the results for the SHC $\sigma^s_H$ in 
the parameter space of 
$\lambda p_F/\varepsilon_F$ and $\tau \varepsilon_F$. Since the SHC
 $\sigma^s_H$ 
is proportional to $(\beta p_{F})^{2}$ and $U/\epsilon_F$
within our calculation scheme, the vertical axis 
is set to be $\sigma^s_H (\beta p_F)^{-2} (\epsilon_{F}/U)$, 
and is normalized by the universal value $e/(8 \pi)$. 
$\sigma^s_H$ has a maximum 
value along the line  $\alpha = 2\lambda p_F \tau \cong 1$ 
at each $\lambda$ and 
the maximum value depends on $\log(\Lambda/p_F)$. 
This conserved spin current is nonzero even for the Rashba model in two dimensions.

 The above results give a hint to look for systems
showing the spin Hall current. One important feature is that the Hamiltonian should involve
$s_z$. Luttinger model \cite{Murakamiscience} satisfies this condition while
the cubic Rashba model does not. Therefore the complete confinement of the electronic motion 
along one direction is not desirable. 
We have assumed that the spin-orbit interaction is unchanged 
by disorder. In reality, 
the impurity potential $v(\mathbf{r})$ induces a
spin-orbit coupling as $(\mathbf{p}\times\nabla v)\cdot \mathbf{s}$.
This is also expected to contribute to the 
extrinsic spin Hall effect.~\cite{Engel} 
This effect is beyond the scope of the present paper.

In conclusion, we derived a general exact formula of
the spin Hall conductivity for the two kinds of 
spin current: (a) the product of spin and velocity operators
and (b) the effective conserved spin current.
The conditions for the non-zero spin Hall current has been clarified and are applied to the Rashba and cubic Rashba models.

The authors thank H.~Fukuyama, B.~I.~Halperin, M.~Onoda, and S.~Y.~Liu 
for stimulating discussions. The work is supported by 
the Grants-in-aid for Scientific Research 
and NAREGI Nanoscience Project from the Ministry of Education, 
Culture, Sports, Science, and Technology.


\begin{thebibliography}{99}
\bibitem{note8pi}
We take the electron charge to be $-|e|=-e$.
\end{thebibliography}
\end{document}